\documentclass[fleqn,usenatbib]{mnras}

\usepackage{newtxtext,newtxmath}
\usepackage{comment}
	
\usepackage[switch, modulo]{lineno}
\usepackage[T1]{fontenc}

\DeclareRobustCommand{\VAN}[3]{#2}
\let\VANthebibliography\thebibliography
\def\thebibliography{\DeclareRobustCommand{\VAN}[3]{##3}\VANthebibliography}

\usepackage{graphicx}	
\usepackage{amsmath}	
\usepackage{newtxtext,newtxmath}

\newcommand{\be}{\begin{equation}} 
\newcommand{\ee}{\end{equation}}
\newcommand{\msun}{\mbox{M$_{\odot}$}}

\usepackage[dvipsnames]{xcolor}

\usepackage{soul}	
\usepackage{xcolor}

\title[The gamma-ray emission of type II-P supernovae]{The first days of type II-P core collapse supernovae in the gamma-ray range}

\author[Cristofari et al.]{P. Cristofari,$^{1}$\thanks{E-mail: pierre.cristofari@obspm.fr}
A. Marcowith,$^{2}$ 
M. Renaud,$^{2}$
V.V. Dwarkadas,$^{3}$
V. Tatischeff~$^{4}$, 
G. Giacinti~$^{5,6,7}$, 
\newauthor
E. Peretti~$^{8}$,
H. Sol~$^{1}$
\\
$^{1}$  LUTH, Observatoire de Paris, CNRS, PSL University, place Jules Jansen, 92190, Meudon, France \\
$^{2}$ Laboratoire Univers et Particules de Montpellier (LUPM), Universit\'e de Montpellier, CNRS/IN2P3, CC72, place Eug\`ene Bataillon, \\ F-34095 Montpellier Cedex 5, France \\
$^{3}$ Department of Astronomy and Astrophysics, University of Chicago, 5640 S Ellis Ave, Chicago, IL 60637, USA \\
$^{4}$ Universit\'e Paris-Saclay, CNRS/IN2P3, IJCLab, 91405 Orsay, France \\
$^{5}$ Max-Planck-Institut f\"ur Kernphysik, Postfach 103980, 69029 Heidelberg, Germany \\
$^{6}$ Tsung-Dao Lee Institute, Shanghai Jiao Tong University, Shanghai 200240, P. R. China \\
$^{7}$ School of Physics and Astronomy, Shanghai Jiao Tong University, Shanghai 200240, P. R. China \\
$^{8}$ Niels Bohr International Academy, Niels Bohr Institute, University of Copenhagen, Blegdamsvej 17, DK-2100 Copenhagen, Denmark \\
}
\date{Accepted XXX. Received YYY; in original form ZZZ}

\pubyear{2015}

\begin{document}
\label{firstpage}
\pagerange{\pageref{firstpage}--\pageref{lastpage}}
\maketitle

\begin{abstract}
Type II-P supernov\ae~(SNe), the most common core-collapse SNe type, result from the explosions of red supergiant stars. Their detection in the radio domain testifies of the presence of relativistic electrons, and shows that they are potentially efficient energetic particle accelerators.
If hadrons can also be accelerated, these energetic particles are expected to interact with the surrounding medium to produce a gamma-ray signal even in the multi--TeV range. The intensity of this signal depends on various factors, but an essential one is the density of the circumstellar medium. Such a signal should however be limited by electron-positron pair production arising from the interaction of the gamma-ray photons with optical photons emitted by the supernova photosphere, which can potentially degrade the gamma-ray signal by over ten orders of magnitude in the first days/weeks following the explosion. 
We calculate the gamma-gamma opacity from a detailed modelling of the time evolution of the forward shock and supernova photosphere, taking a full account of the non-isotropy of the photon interactions. We discuss the time-dependent gamma-ray TeV emission from type II-P SNe as a function of the stellar progenitor radius and mass-loss rate, as well as the explosion energy and mass of the ejected material. 
We evaluate the detectability of the SNe with the next generation of Cherenkov telescopes. We find that, while most extragalactic events may be undetectable, type II-P SNe exploding in our Galaxy or in the Magellanic Clouds should be detected by gamma-ray observatories such as the upcoming Cherenkov Telescope Array.
\end{abstract}

\begin{keywords}
Stars: supernovae: general – Interstellar medium: Cosmic Rays – gamma-rays: general.
\end{keywords}


\section{Introduction}
Core collapse supernovae (CCSNe) result from the explosive deaths of massive stars with masses $\gtrsim 10 \msun$ \citep{2003ApJ...591..288H}. The explosion produces a fast shock propagating out into the circumstellar medium (CSM), and a second 'reverse' shock expanding back into the ejecta. SNe are expected to radiate over the entire multi-wavelength spectrum, and for now have been observed at all except at the very highest energies.  In the very--high--energy (VHE) range, CCSNe are important because they have been invoked as being capable of accelerating particles up to, or even above, the PeV
range~\citep[see
  e.g.,][]{tatischeff2009,bell2013,schure2014,marcowith2014,murase2014,cardillo2015,Giacinti15,
  zirakashvili2016,petropoulou17, bykov2918,marcowith2018,
  murase2019,fang2019} and {reviews by \citet{bykov18,
    tamborra18}}.

Besides, the PeV range is an important milestone in cosmic ray (CR) physics. The spectrum of CRs measured at the Earth follows a remarkable power law in energy, with mild deviation, up to the \textit{knee} domain, at $\sim$ 1--3 PeV, where a major spectral deviation occurs. The sources producing the bulk of CRs are expected to accelerate particles up to the PeV range, and to lie within the Galaxy~\citep{cristofari2020b}. The hunt for pevatrons is now a well--identified key target of the astroparticle community~\citep[see
  e.g.,][for reviews on the topic]{blasi2019,gabici2019,CTAscience,cristofari_review}.

The possibility that CCSNe could accelerate PeV particles is a strong motivation for their study. The acceleration of PeV particles should directly lead to the production of a potentially detectable amount of gamma rays from the GeV to the multi--TeV range~\citep{kirk1995}, due to the interaction of accelerated protons with interstellar medium (ISM) nuclei, through the production of pions. This possibility has been discussed in various works \citep{dwarkadas2013,marcowith2018, murase2019, fang2019}, and recent studies based on {\it Fermi}-LAT data have reported marginally-to-moderately significant variable high-energy emission towards the peculiar H-rich superluminous SN iPTF14hls~\citep{Yuan18}, the nearby type IIP event SN 2004dj~\citep{Xi20}, and in the direction of the SN candidates AT2019bvr and AT2018iwp, with a flux increase within six months after the discovery date~\citep{Prokhorov21}.
Future instruments optimized in the TeV and multi--TeV range~\citep{CTAscience} might then be capable of detecting extragalactic CCSNe.

A primary limiting factor for the detection of CCSNe in the TeV range is the distance, which determines the flux reaching the observer. The flux is further expected to decrease inversely with time \citep{tatischeff2009}. The $\gamma$-ray signal should also be degraded by the interaction of gamma-rays with the low energy photons from the SN photosphere~\citep{gould1966,gould1967} during the first stages of the stellar outer envelope expansion, resulting in the production of electron--positron pairs. This two--photon annihilation process has been discussed in various astrophysical contexts, generally under the assumption of isotropy~\citep{aharonian2008,tatischeff2009,murase2014,wang2019}. 
None of these calculations took into account the anisotropy inherent in the problem, or the time of flight of the photons from the photosphere. The calculations carried out by \citet{cristofari2020}, and outlined below, include some important improvements that considerably increase the accuracy, and enhance the validity, of the solutions: the time retardation effects of the SN photosphere, the Doppler effect over the frequency space, the full anisotropic calculation of the two-photon annihilation process, and a detailed modelling of the time evolution of both the forward shock and the SN photosphere. This more careful treatment of the problem, which takes into account the diverging evolution of the radius of the SN photosphere and the forward shock, can produce results that substantially differ from the isotropic assumption, as illustrated in \citet{cristofari2020} in the case of SN 1993J. 

SN~1993J was discovered and well-monitored from the optical to radio range since the first days after its explosion in the Galaxy M81 (NGC3031)~\citep{ripero1993}. It was classified as a type IIb SN~\citep{maund2004} due to the initial detection of H in the spectrum that later disappeared. The progenitor star was found to be in the range 13--20 M$_{\odot}$~\citep{vandyk2005,Marcaide2009}. The close distance, and the high inferred mass-loss rate, makes it an appealing candidate for the detection of an extragalactic SN in the gamma-ray range. 
  
Type IIb SNe occur at a relatively low rate, comprising $\lesssim 5$ \% of the total core-collapse SNe \citep{smartt2009}. The bulk of CCSNe expand in a lower density surrounding medium. The largest fraction of CCSNe are the type II-P SNe, accounting for typically $\sim 40-50$\% of CCSNe. Here the "P" stands for plateau, since these SNe show a plateau in their optical light curve lasting for several months. The high frequency means that the likelihood of a Type II-P SN being detected in an optical survey of nearby SNe is much higher. In this paper, we therefore focus our attention on the expected gamma-ray signal from Type II-P SNe in the $\sim 30$ days following the explosion of the SN where the unabsorbed flux is expected to be the highest \citep{tatischeff2009, marcowith2018}.

As we will show in this paper, the predicted flux is very sensitive to the SN parameters. Those affecting the gamma-ray flux can vary over a wide range, which can lead to large variations in the flux itself. It is therefore important to study the expected flux over a plausible range of parameters. We explore the parameter space of SN quantities that can possibly affect the gamma-ray signal: the total explosion energy of the SN, the ejecta mass, the mass-loss rate of the wind, and radius and temperature of the progenitor star. Their individual and collective effects on the absorption of the gamma-ray signal are investigated. The calculation of the time-dependent opacity is carried out in a manner analogous to the one presented in~\citet{cristofari2020}.



The lay-out of the paper is as follows: in Sec.~\ref{sec:evolution} we describe the dynamics of the SN photosphere and SN shock, in Sec.~\ref{sec:gamma} we describe the calculation of the gamma-ray signal including the gamma-ray attenuation by the two-photon annihilation process, in Sec.~\ref{sec:results} we present our results as a function of the parameters describing the stellar progenitor and the SN explosion, and we conclude in Sec.~\ref{sec:conclusions}. 

\section{Evolution of the supernova shock and photosphere}
\label{sec:evolution}

The evolution of the photospheric radius and the radius of the forward shock in the first few days after explosion can be described using approximate analytical expressions. Before the SN explosion, the progenitor star of a II-P SN is a red supergiant (RSG), in which the pre-explosion density profile within the shell, under the assumption of a efficiently convective envelope, can be approximated by~\citep{chandrasekhar1939}:
\begin{equation}
\rho_0(r_0) = \rho_{1/2} (1 - r_0/R_{\star})^{n_p}~,
\end{equation}
where $n_p=3/2$ and $R_{\star}$ is the radius of the star, and  $\rho_{1/2}$ the density at $r_0=0$. After the SN explosion, the photospheric radius can be written in Gaussian cgs units as \citep{rabinak2011}:
\begin{equation}
\label{eq:rph}
\begin{aligned}
r_{\rm ph}(t) = 2.9 \times 10^{14}~\text{cm} \left( \frac{f}{0.1} \right)^{-0.062} \left( \frac{E_{\rm SN}}{10^{51}~\text{erg}}\right)^{0.41}  \\
 \times  \left( \frac{\kappa}{0.34~\text{cm}^2\text{/g}} \right)^{0.093} \left( \frac{M_{\rm ej}}{M_{\odot}}\right)^{-0.31} \left( \frac{t}{\text{days}} \right)^{0.81} 
 \end{aligned}
\end{equation}
and the photospheric temperature: 
\begin{equation}
\label{eq:Tph}
\begin{aligned}
T_{\rm ph}(t) = 1.7~\text{eV} \left( \frac{f}{0.1} \right)^{-0.037} \left( \frac{E_{\rm SN}}{10^{51}~\text{erg}}\right)^{0.027}    \left( \frac{R_{\star}}{10^{13}~\text{cm}}\right)^{1/4}  \\
\times  \left( \frac{\kappa}{0.34~\text{cm}^2\text{/g}} \right)^{-0.28} \left( \frac{M_{\rm ej}}{M_{\odot}}\right)^{-0.054}    \left( \frac{t}{\text{days}} \right)^{-0.45} 
\end{aligned}
\end{equation}
where $f= \rho_{1/2}/\bar{\rho_0}$, with $\bar{\rho_0}= 3 M_{\rm ej}/(4 \pi R_{\star}^3)$  the  average ejecta density. Typical values for $f$ are found for RSG in the range [0.079,0.13], $M_{\rm ej}$ is the ejecta mass, $E_{\rm SN}$ the total explosion energy (kinetic energy), and $\kappa$ the opacity, assumed to be time-and-space-independent (e.g., the opacity is dominated by Thomson scattering with constant ionization). We note that the dependency of both $r_{\rm ph}$ and $T_{\rm ph}$ on $f$ is weak. $r_{\rm ph}$ is dominantly dependent on $E_{\rm SN}$ and $M_{\rm ej}$. On the other hand, $T_{\rm ph}$ mostly depends on $R_{\star}$ and $\kappa$.

\begin{figure}
\includegraphics[width=.48\textwidth]{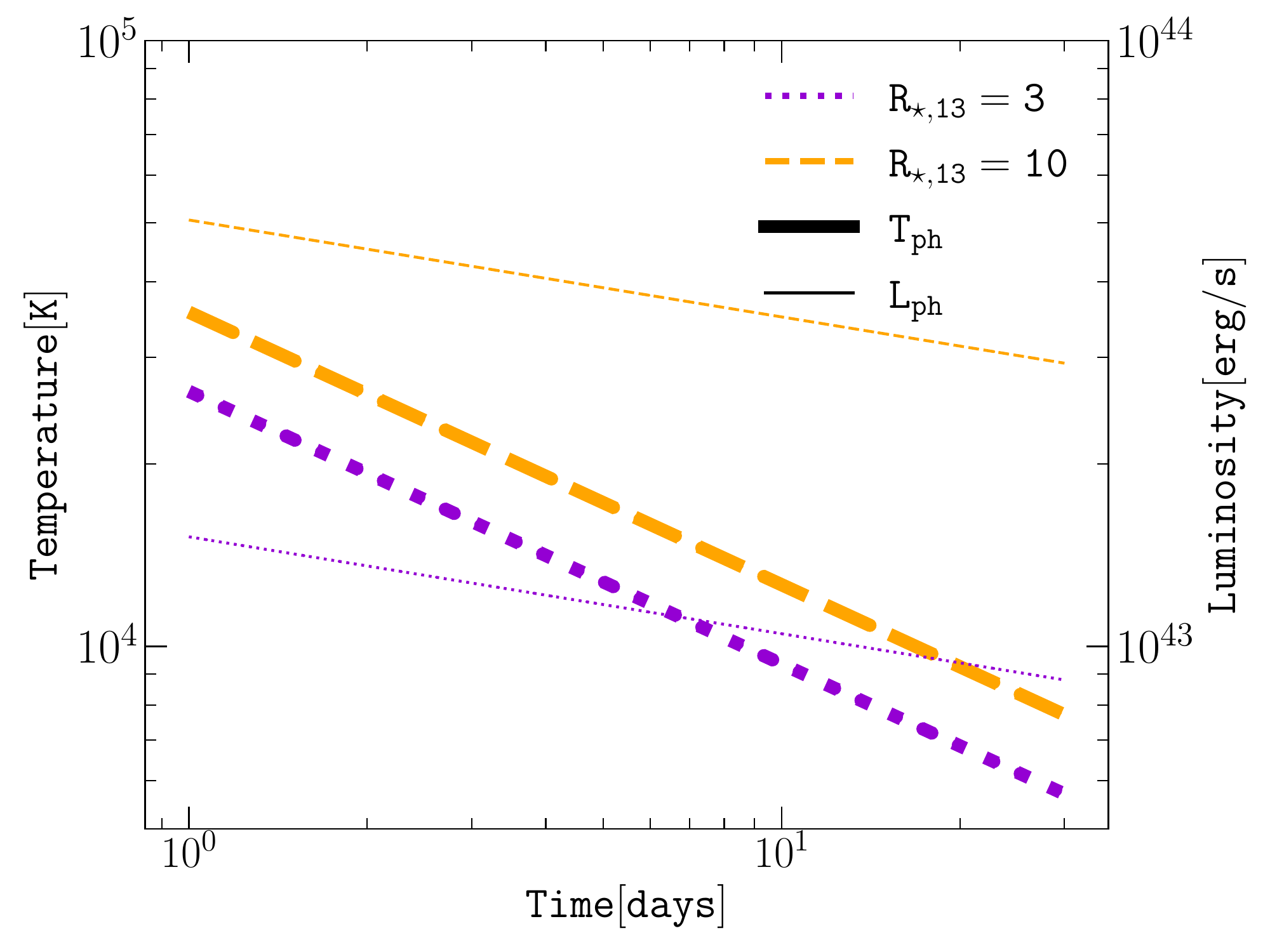}
\caption{Time evolution of the photospheric temperature (thick lines) and luminosity (thin lines) in the case of $R_{\star}=3 \times 10^{13}$ cm (violet dotted) and $R_{\star}=10^{14}$ cm (orange dashed).$E_{\rm SN}=10^{51}$ erg, $M_{\rm ej}=4$ M$_{\odot}$.}
\label{fig:temperature}
\end{figure}

Fig.~\ref{fig:temperature} shows the evolution of the photospheric temperature and luminosity. The latter quantity is calculated as $L_{\rm ph} = 4 \pi r_{\rm ph}^2 \sigma_{\rm SB} T_{\rm ph}^4$, where $\sigma_{\rm SB}$ is the Stefan-Boltzmann constant.

\begin{figure}
\includegraphics[width=.5\textwidth]{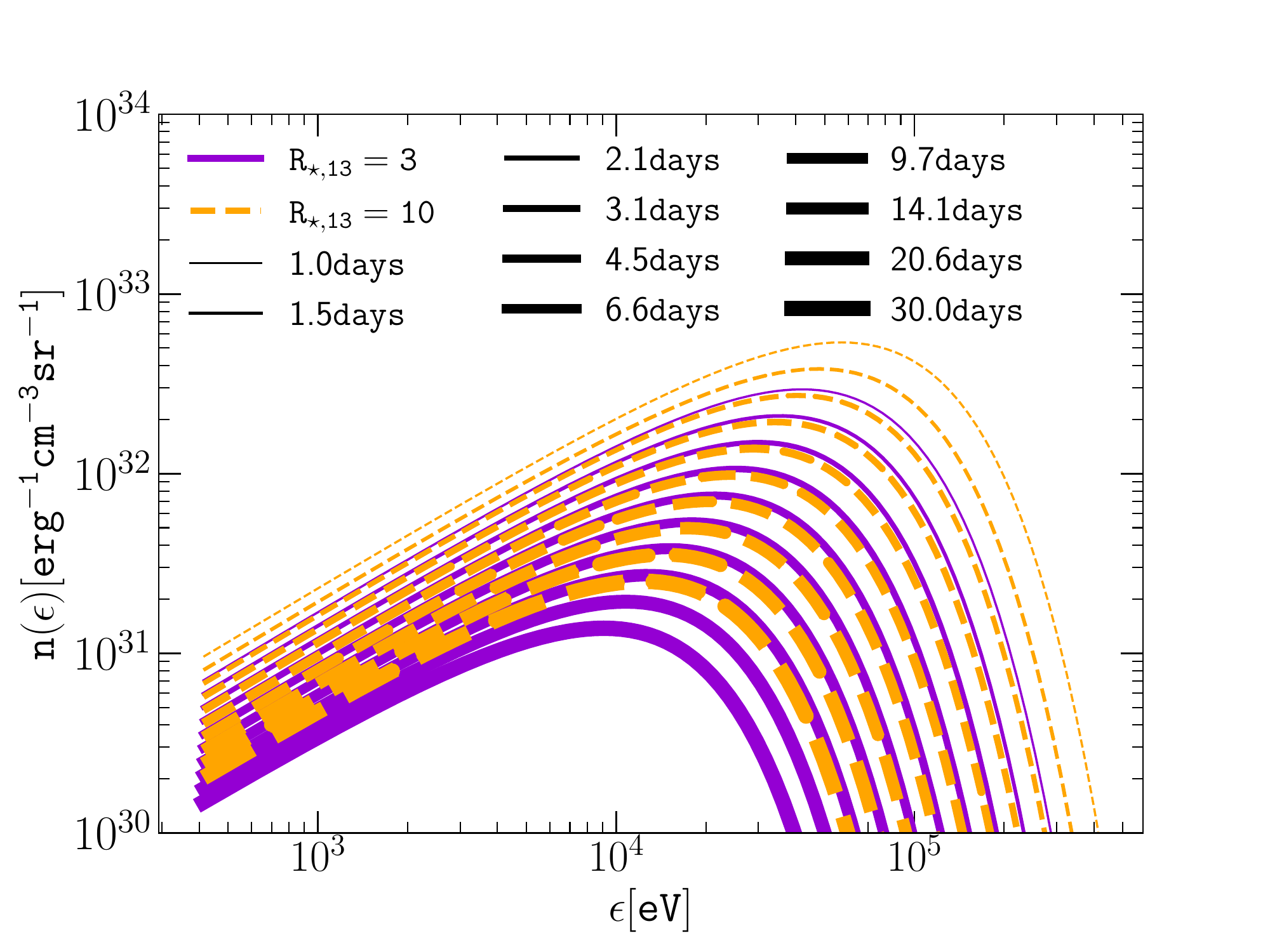}
\caption{Spectral energy distribution of the low energy photons (Eq.~\eqref{eq:n_e}) emitted by the photosphere for $R_{\star}=3 \times 10^{13}$~cm (violet solid) and $R_{\star}=10^{14}$~cm (orange dashed). The increased thickness follows the time evolution.}
\label{fig:photons}
\end{figure}

The photospheric photon density is assumed to follow a blackbody distribution with a time-dependent temperature $T_{\rm ph}(t)$:
\begin{equation}
\label{eq:n_e}
    n (\epsilon, t)= \frac{2 \epsilon^2}{h^3 c^3} \frac{1}{\exp (\epsilon/kT_{\rm ph}(t))-1} ~~\text{cm}^{-3} ~\text{erg}^{-1} ~\text{sr}^{-1}.
\end{equation}
Figure~\ref{fig:photons} shows the time evolution of $n (\epsilon, t)$ during the first 30 days after the SN explosion, for two values of the progenitor star radius.


Only the acceleration of particles at the forward shock is considered here, so that the effects of the reverse shock are not presently included~\citep{ellison2005,telezhinsky2012,leahy2017}. Indeed, the reverse shock is typically propagating through higher densities, with a lower velocity so that the maximum energy is lower~\citep{telezhinsky2012}. In case particle acceleration at the reverse shock would produce gamma rays: the lower velocity of the shock tends to reduce the maximum energy of accelerated particles, but the higher density in which the shock is propagating can lead to an enhanced production of gamma rays but also stronger losses through pp collisions. Therefore, all in all any acceleration at the reverse shock would increase the total gamma--ray signal and this makes our approach a lower limit on the estimate of gamma rays from type II-P. 


The SN strong forward shock resulting from the SN explosion can be described by self--similar solutions. We adopt the description proposed in~\citet{tang2017} for a SN exploding in a wind profile $\rho_{\rm w} (r) = \eta_{s} r^{-s}$, with $\eta_{s}=\dot{M}_{\rm w}/(4 \mu \pi v_{\rm w})$, where $\dot{M}_{\rm w}$ and $v_{\rm w}$ are the mass-loss rate and wind terminal velocity of the RSG progenitor, and $\mu=(1+4 f_{\rm He})/(1+f_{\rm He})$ accounts for a fraction $f_{\rm He}=10$\% of He in the ISM. The time evolution of the shock radius is then of the form:
\begin{equation}
    R_{\rm sh} (t)= R_{\rm ch} \left[ \left( \zeta \left(\frac{t}{t_{\rm ch}}\right)^{(n-3)/(n-s)} \right)^{-\alpha} + \left( \xi  \left(\frac{t}{t_{\rm ch}}\right)^2 \right)^{-\alpha/(5-s)} \right]^{-1/\alpha}
    \label{eq:Rsh}
\end{equation}
where $R_{\rm ch}= M_{\rm ej}^{1/(3-s)} \eta_{s}^{-1/(3-s)}$ and $t_{\rm ch}= E_{\rm SN}^{-1/2} M_{\rm ej}^{\frac{5-s}{2(3-s)}}
\eta_{s}^{-1/(3-s)}$, and the parameters $\zeta$, $\xi$, and $\alpha$ depend on the SN type.
In a wind density profile, the characteristic radius and time read:  
\begin{equation}
R_{\rm ch } \approx 129 \left(\frac{M_{\rm ej}}{M_{\odot}} \right) \left ( \frac{M_{\rm w}}{10^{-6} \text{M}_{\odot}/\text{yr}}\right)^{-1} \left( \frac{v_{\rm w}}{10^6 \; \text{cm/s}}\right) \text{pc}
\end{equation}

\begin{equation}
\begin{aligned}
t_{\rm ch } \approx 17.7  \left(\frac{E_{\rm SN}}{10^{51} \text{erg}} \right)^{-1/2} \left( \frac{M_{\rm ej}}{M_{\odot}} \right)^{3/2}\\
\times \left ( \frac{M_{\rm w}}{10^{-6} \text{M}_{\odot}/\text{yr}}\right)^{-1} \left( \frac{v_{\rm w}}{10^6 \; \text{cm/s}}\right) \text{kyr}
\end{aligned}
\end{equation}

 Eq.~\eqref{eq:Rsh} describes the shock evolution provided that $n>5$, with $n$ the slope of the density profile of the envelope of the exploding star. For a typical type II-P SNe, $n=10$, $\zeta=1.03$, $\xi=0.477$, $s=2$ and $\alpha=4.56$. The corresponding shock velocity is given by the derivative $v_{\rm sh} = \text{d}R_{\rm sh}/\text{d}t$.

Hence, Eq.~\eqref{eq:Rsh} can be reformulated explicitly with physical parameters of interest for type II-P SNe: 
\begin{equation}
\label{eq:Rsh2}
\begin{aligned}
 R_{\rm sh} (t)\approx 4.2 \; 10^{14}~\text{cm} \left( \frac{E_{\rm SN}}{10^{51}~\text{erg}} \right)^{0.44} \left( \frac{\dot{M}_{\rm w}}{10^{-6}~ \text{M}_{\odot}\text{/yr}}\right)^{-0.125} \\
 \times \left( \frac{v_{\rm w}}{10^6~\text{cm/s}}\right)^{0.125} \left(\frac{M_{\rm ej}}{\text{M}_{\odot}} \right)^{-0.31} \left(\frac{t}{\text{days}}\right)^{0.875} 
 \end{aligned}
\end{equation}

Deviations from the value $n=10$ have been proposed for outer SN ejecta in RSGs, such as~\citet{mat99} fixing it at $n=11.73$. 
 
\begin{figure}
\includegraphics[width=.48\textwidth]{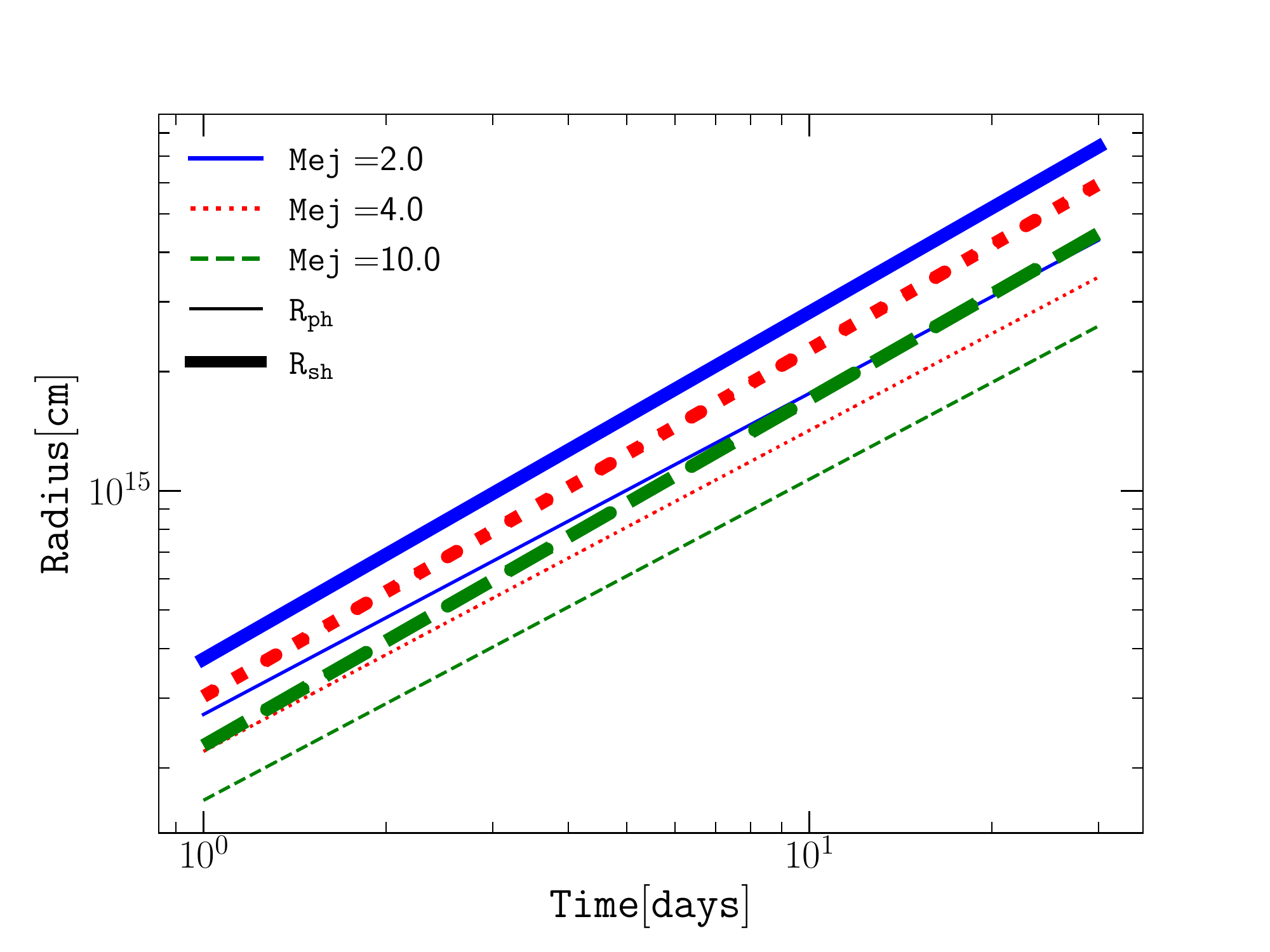}
\caption{Time evolution of the shock (thick) and photospheric (thin) radii, for an ejecta mass $M_{\rm ej}=2,4,10$ M$_{\odot}$ (blue solid, red dotted, and green dashed, respectively), and a mass-loss rate $\dot{M}_{\rm w}=10^{-6}$ M$_{\odot}$/yr.}
\label{fig:radius}
\end{figure}

We illustrate in Fig.~\ref{fig:radius} the evolution of $r_{\rm ph}$ and $R_{\rm sh}$ during the first 30 days after the SN explosion, for typical values of the mass-loss rate $\dot{M}_{\rm w}=10^{-6}$~M$_{\odot}$/yr and ejecta mass $M_{\rm ej}=2,4,$ and $10$ M$_{\odot}$.

\section{gamma-ray emission}
\label{sec:gamma}
\subsection{Unattenuated gamma-ray flux}
Our work focuses on describing the effects of the gamma-gamma attenuation, which is the dominant effect on the gamma--ray fluxes from a SN shock in the first days/weeks after the SN explosion. We rely on a simple analytic description of the unattenuated gamma-ray signal. A reasonable estimate of the unattenuated gamma-ray flux reads:
\begin{equation}
F_{\gamma} (> 1~\text{TeV}) \approx \frac{q_{\gamma} (>1 \text{TeV}) \bar{n} \bar{\epsilon} V}{4 \pi D^2} 
\end{equation}
with $q_{\gamma} (>1~\text{TeV}) \approx 10^{-17}$ photons s$^{-1}$ erg$^{-1}$ cm$^{3}$ is the gamma-ray emissivity,  which depends on the spectrum of particle accelerated at the shock~\citep{drury1994}, $\bar{n}$ is the mean ISM density in the volume $V$ downstream occupied by the CRs, $D$ is the source distance, and $\bar{\epsilon} \approx 3 P_{\rm CR}$ is the mean energy
density of CRs, $P_{\rm CR}$ being the CR pressure at the shock front.. The volume occupied by CRs downstream of the shock is typically $V \approx \frac{4\pi}{3} \left(r_{\rm sh}^3 - r_{\rm CD}^3 \right) \approx \frac{2 \pi}{3} r_{\rm sh}^3$, where the contact discontinuity is at $r_{\rm CD} =r_{\rm sh}/1.24$ for $n=10$, according to the self-similar hydrodynamic  model~\citep{chevalier1982}. For a strong shock with a compression factor $r=4$, the spectrum of accelerated particles in the test-particle limit follows at high energy a power-law in energy  $\propto E^{-2}$. However, several effects can potentially harden~\citep[see e.g.,][for non-linear DSA]{malkov2001,amato2006}, or steepen~\citep[see e.g.,][for discussions on pre- or post-cursor effects]{Zirakashvili12,haggerty2020,caprioli2020} the spectrum of accelerated particles. As a matter of simplicity we do not enter in such details in this work. A model including non-linear CR and magnetic back reaction over the shock solutions will be presented elsewhere. 

The typical gamma-ray flux at the SN shock thus reads: 
\begin{equation}
\begin{aligned}
F_{\gamma} (> 1\text{TeV}) \approx 2.5 \times 10^{-14}~\text{cm}^{-2}~\text{s}^{-1}~\left( \frac{\xi_{\rm CR}}{0.1}\right) \left( \frac{\dot{M}_{\rm w}}{10^{-6}~ \text{M}_{\odot}/\text{yr}} \right)^2   \\
\times \left( \frac{v_{\rm w}}{10^{6}~\text{cm/s}}\right)^{-2} \left( \frac{D}{\text{Mpc}} \right)^{-2} \left( \frac{v_{\rm sh}(t) }{10^9 ~\text{cm/s}}\right)^2 \left( \frac{R_{\rm sh}(t)}{10^{14}~\text{cm}}\right)^{-1} ,
\end{aligned}
\label{eq:gamma_flux}
\end{equation}
where $\xi_{\rm CR}=P_{\rm CR}/\rho_{\rm w}v_{\rm sh}^2$ is the CR pressure normalised to the kinetic pressure of the shock. 
Finally, we assume that the typical maximum energy reached at type II-P CCSNe is suffiently high $\gtrsim 100$ TeV - 1 PeV  in the month after the SN explosion~\citep{bell2013,schure2014,Inoue21}, so that the signal in the gamma--ray domain of interest is not affected.


\begin{figure}
\includegraphics[width=.45\textwidth]{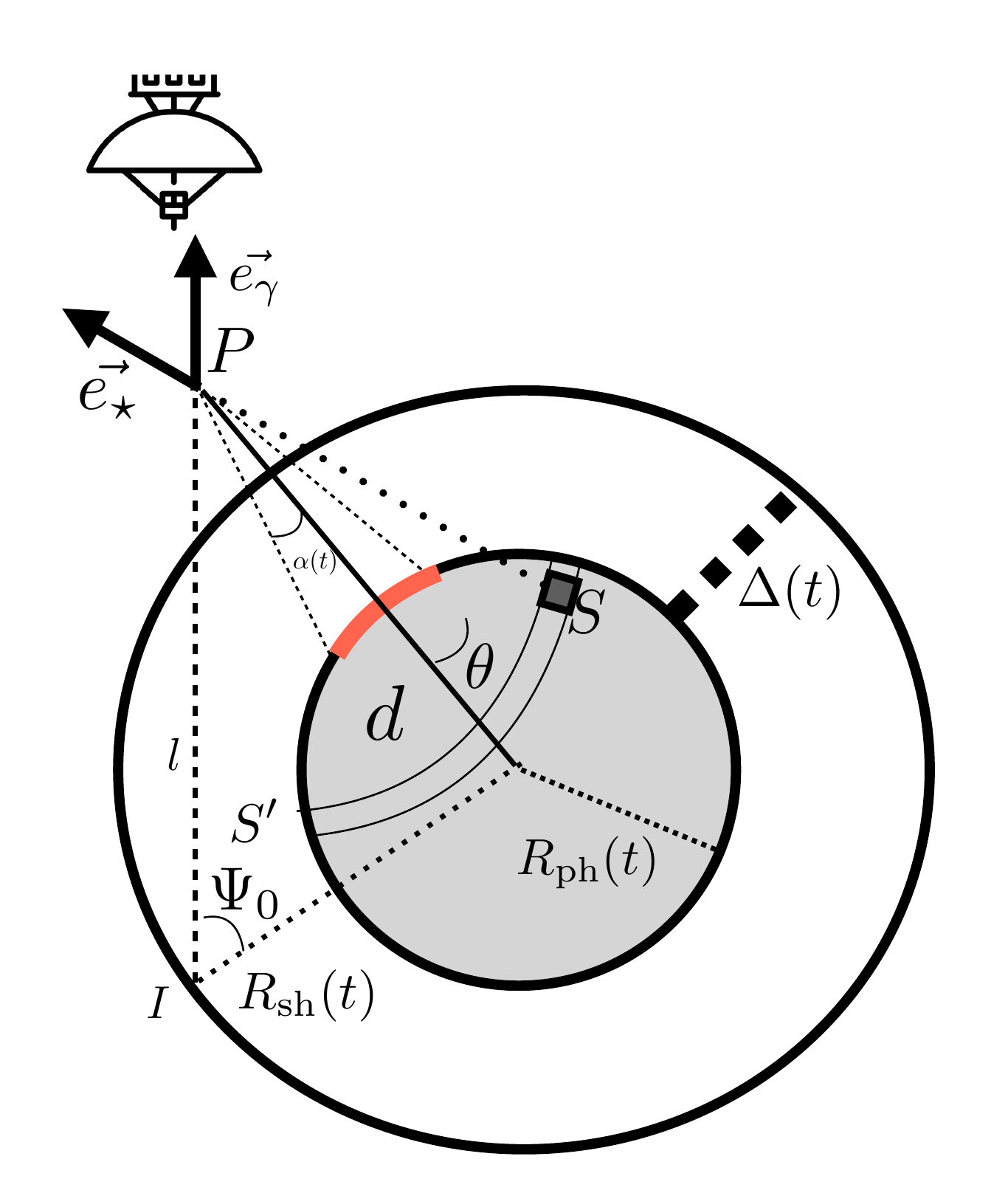}
\caption{2D diagram of the physical problem. The photosphere (grey
  inner disk) emits photons at (S) which interacts at (P) with gamma-ray photons emitted at (I). The area marked by the thick red line represents the region of origin of low-energy photons which can interact with a gamma-ray emitted at (I) (see text). The red (thick) region of the photosphere illustrates that only a fraction of the photosphere, emitting photons at times $t_1$ will interact with gamma rays in P at a time $t_0$.
  }
\label{fig:schema}
\end{figure}

\subsection{Pair production gamma-ray attenuated flux}
The calculation of the absorption due to pair-production is performed as in~\citet{cristofari2020}. A 2D representation of the geometrical problem is shown in Fig.~\ref{fig:schema}. To compute the total photon-photon optical depth at a given gamma--ray energy $E$, one needs to do an integration over six quantities: $\epsilon$, the soft photon energy; $\theta$ the polar angle between the direction of the center of the photosphere and the location of the soft photon emitting region, as view from the interaction point (P); $\phi$, the corresponding azimuthal angle; $l$, the distance between (I) the gamma-ray emitting region and (P) ; $\psi_0$, the angle of the emitted gamma-ray photon at the interaction point relative to the radial direction; and $t$, the time after the SN explosion:
\begin{equation}
\label{eq:tau_general}
\begin{aligned}
\tau_{\gamma \gamma}(E)=& \int_{0}^{t} \text{d}t' \int_{\psi_{0, \rm min}}^\pi \text{d}\psi_0 \int_{0}^{+\infty} \text{d}l  \\ 
& \int_{c_{\rm min}}^{1} \text{d} \cos \theta \int_{0}^{2 \pi}\text{d}\phi \int_{\epsilon_{\rm min}}^{+\infty} \text{d}\epsilon \frac{\text{d}\tau_{\gamma \gamma}}{\text{d}\epsilon \text{d}\Omega \text{d}l} \ ,
\end{aligned}
\end{equation}
where the differential absorption opacity reads~\citep{gould1966,gould1967}: 
\begin{equation}
\text{d}\tau_{\gamma\gamma}= \left(1- {\bf e_{\gamma} e_{\star}} \right)n_{\epsilon} \sigma_{\gamma \gamma} \text{d}\epsilon \text{d}\Omega \text{d}l~, 
\end{equation}
with ${\bf e_{\gamma}}$ and ${\bf e_{\star}}$ denoting the direction of the interacting gamma-ray photon and soft photon respectively. The cross section $\sigma_{\gamma \gamma}$ for the pair production process $\gamma + \gamma \rightarrow e^+ + e^-$ is derived in~\citet{gould1967}, $\text{d}\Omega= \sin \theta \text{d}\phi \text{d}\theta$ is the solid angle of the surface emitting the photons of energy $\epsilon$ and $n_{\epsilon}$ is the radiation density. Eq.~(\ref{eq:tau_general}) is a generalization of Eq.~(A.8) of~\citet{dubus2006} which takes into account temporal effects. It requires the calculation of two more integrals on the time $t$ and emission angle $\psi_0$.

The gamma-ray photon flux can also be degraded by electron-positron production in the Bethe-Heitler process through their interaction with ambient nuclei \citep{murase2014}. As the incident gamma-ray photons are isotropically distributed the opacity due to the Bethe-Heitler process reads:
\begin{equation}
    \tau_{\rm BH} = R_{\rm int} n_{\rm i} \sigma_{\rm BH} \ ,
\end{equation}
where the Bethe-Heitler cross section for photons with energies $\gg m_{\rm e} c^2$ can be expressed as 
$\sigma_{\rm BH} \sim 2.3~10^{-27} \left(\ln(E_{\rm ph,\gamma, GeV})+5.7\right)$ (we have assumed an effective charge $Z = 1.14$ to account for the presence of He). The typical radius over which the interaction occurs is $R_{\rm int} = \Lambda R_{\rm sh}$ with typically $\Lambda \sim 1$. We find:
\begin{equation}
\begin{aligned}
    \tau_{\rm BH} \simeq 6~10^{-5} \Lambda \left( \frac{\dot{M}_{\rm w}}{10^{-6}~ \text{M}_{\odot}/\text{yr}} \right) \\
    \times \left(\frac{R_{\rm sh}(t)}{10^{14} \text{cm}} \right)^{-1} \left( \frac{v_{\rm w}}{10 \text{km/s}}\right)^{-1}   \left(\ln(E_{\rm ph,\gamma, GeV})+5.7 \right) \ .
    \end{aligned}
\end{equation}
The latter is usually $\ll 1$, unless the mass--loss rate exceeds a few times $10^{-2} M_\odot/\rm{yr}$, then GeV photons and beyond can be absorbed. In type II-P SNe, we thus do not expect this process to affect substantially the gamma--ray flux.

\section{Results and discussion}
\label{sec:results}
The model for the gamma-ray emission of a typical type II-P CCSN adopted here depends on a few physical parameters: the total explosion energy $E_{\rm SN}$, the mass-loss rate of the pre--SN wind $\dot{M}_{\rm w}$, the wind terminal velocity $v_w$, the ejecta mass M$_{\rm ej}$ and the radius of the progenitor star $R_{\star}$. The influence of these parameters on the gamma-ray signal is discussed next. 

Red supergiant (RSG) stars can have mass-loss rates from about 10$^{-8} \msun$ yr$^{-1}$ to 10$^{-4} \msun$ yr$^{-1}$ \citep{mj11}. However, it had often been noted that the progenitors of II-Ps appeared to arise from RSG stars that were lower than about 20 $\msun$. This was first quantified by \citet{smartt2009} using optically identified SNe. \citet{Dwarkadas2014} reached a similar conclusion using the X-ray lightcurves of SNe. A larger dataset allowed \citet{smartt2015} to consolidate their early arguments. In general it appears that while RSG stars may have higher mass-loss rates, Type II-P SNe appear to arise from the lower end of the RSG mass function. And since the mass-loss rates are directly proportional to the RSG mass \citep{mj11}, the II-Ps generally expand in a medium with lower mass-loss rates. Therefore, we have assumed that II-P SNe expand in a medium with mass-loss rates $\dot{M}_{\rm w}=10^{-8}, 10^{-7}, 10^{-6}$ and $10^{-5}$ M$_{\odot}$/yr respectively. We discuss the possibility of higher mass loss rates in section \ref{sec:conclusions}. 
\begin{figure}
\includegraphics[width=.48\textwidth]{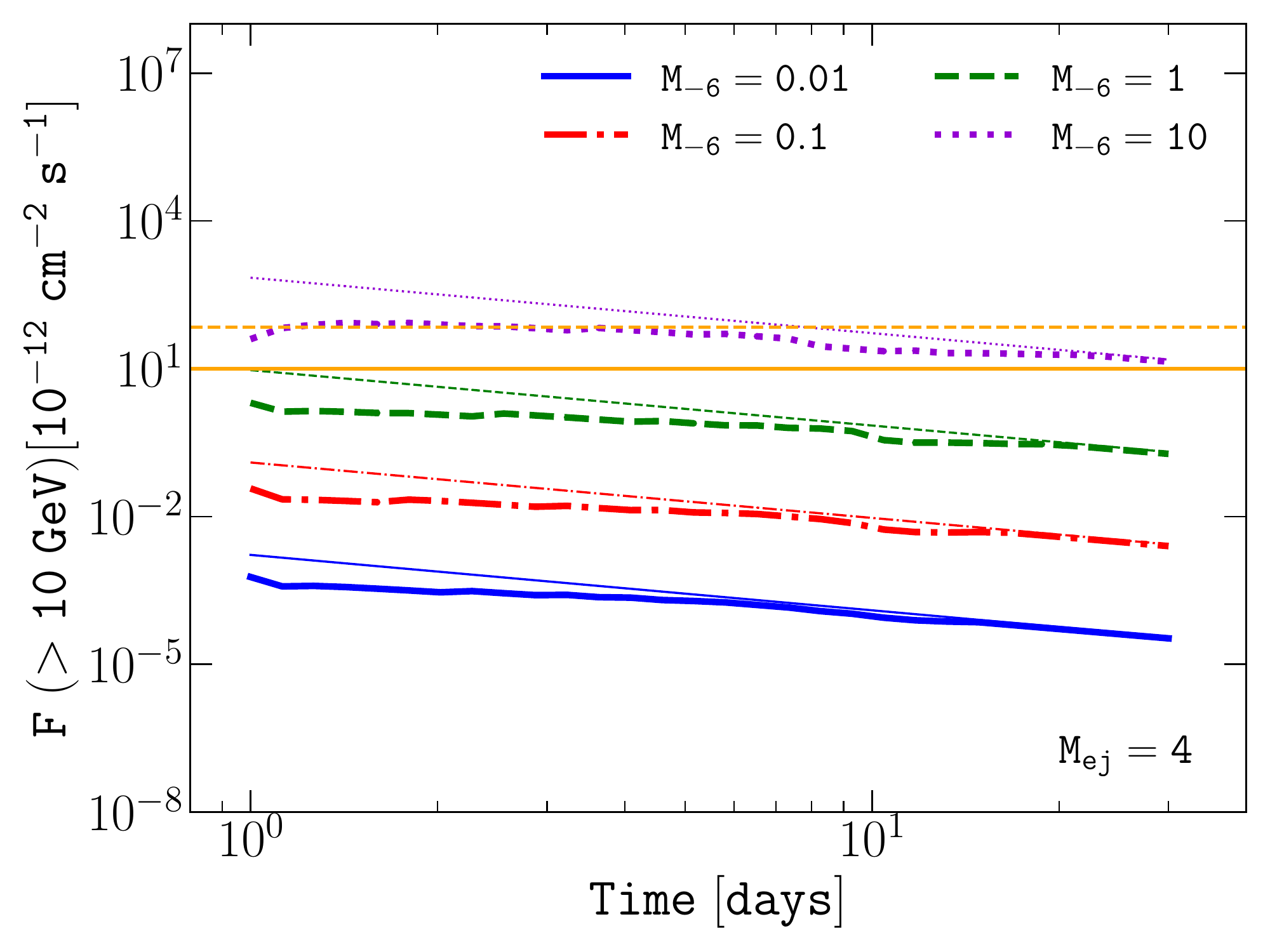}\\
\includegraphics[width=.48\textwidth]{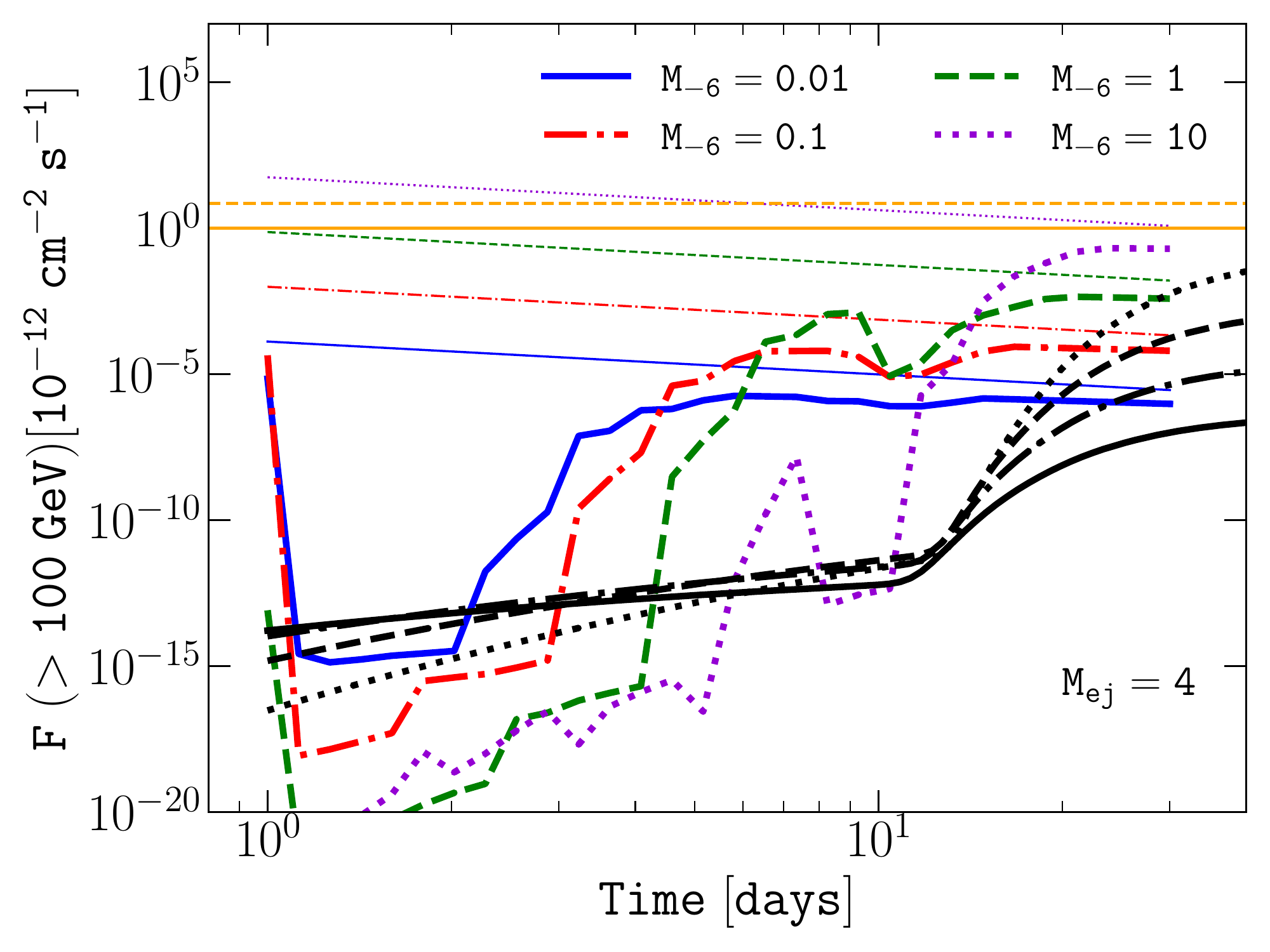}
\includegraphics[width=.48\textwidth]{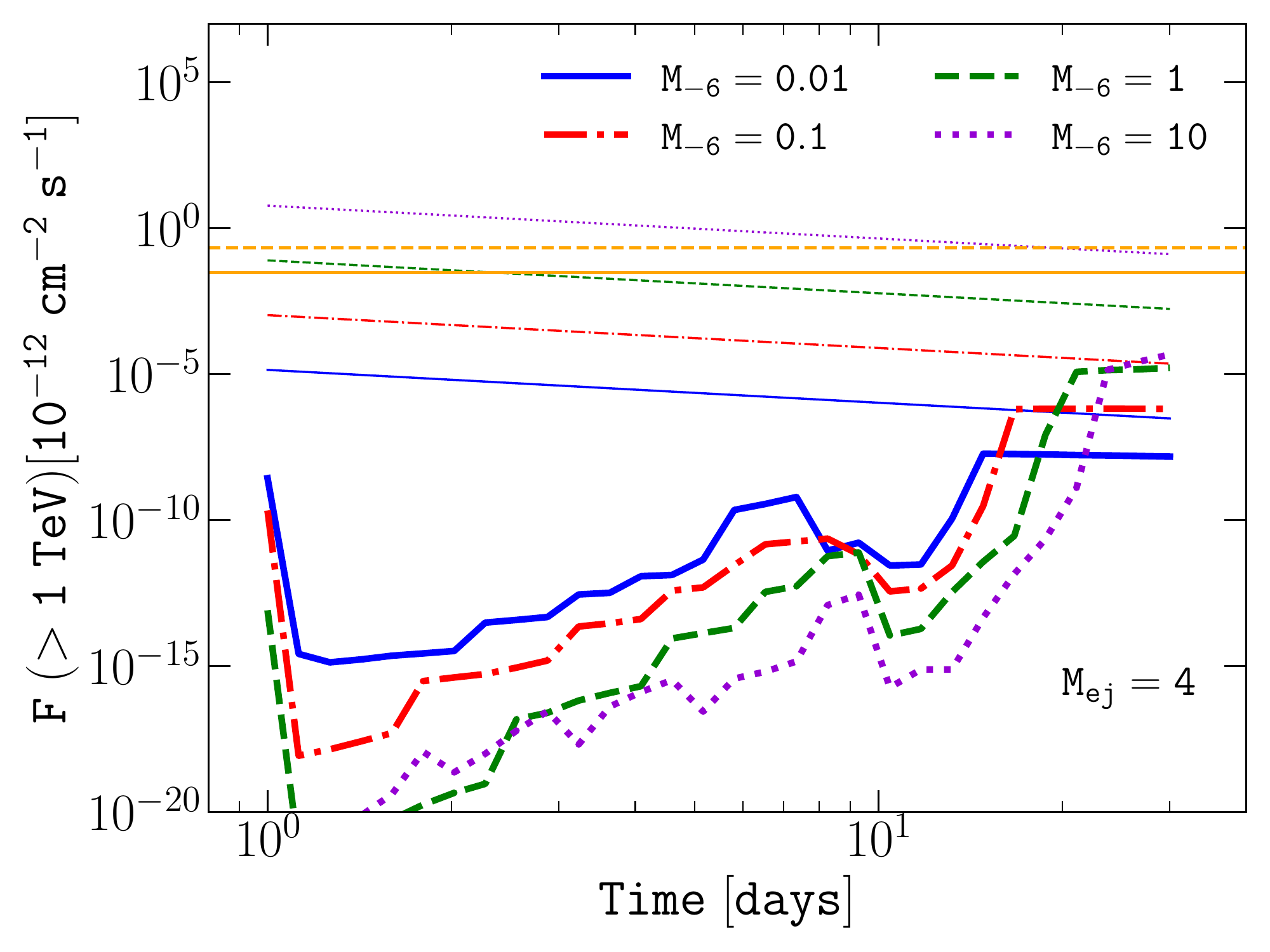} \\ 
\caption{ Time evolution of the integrated photon flux above 10 GeV (top), 100 GeV (middle) and 1 TeV (bottom). The source distance is $D=1$~Mpc, the ejecta mass is $M_{\rm ej}=4$ M$_{\odot}$, the progenitor radius is $R_{\star}=3 \times 10^{13}$~cm. The mass-loss rate of the wind is $\dot{M}_{\rm w}=10^{-8}, 10^{-7}, 10^{-6}$ and $10^{-5}$ $M_{\odot}$/yr, is shown as solid (blue), dot-dashed (red), dashed (green) and dotted (violet) lines, respectively. The corresponding unattenuated fluxes are shown as thin lines. The typical sensitivity of CTA  for 50 hours (solid orange  horizontal line) and 2 hours (dashed orange  horizontal line) is shown as a guiding-eye for the reader~\citep{fioretti2016}. On the middle panel we additionally plot (black lines) the corresponding gamma-ray fluxes obtained under the approximation of an homogeneous source as in~\citet{wang2019}.}
\label{fig:Mej4}
\end{figure}

\begin{figure}
\includegraphics[width=.48\textwidth]{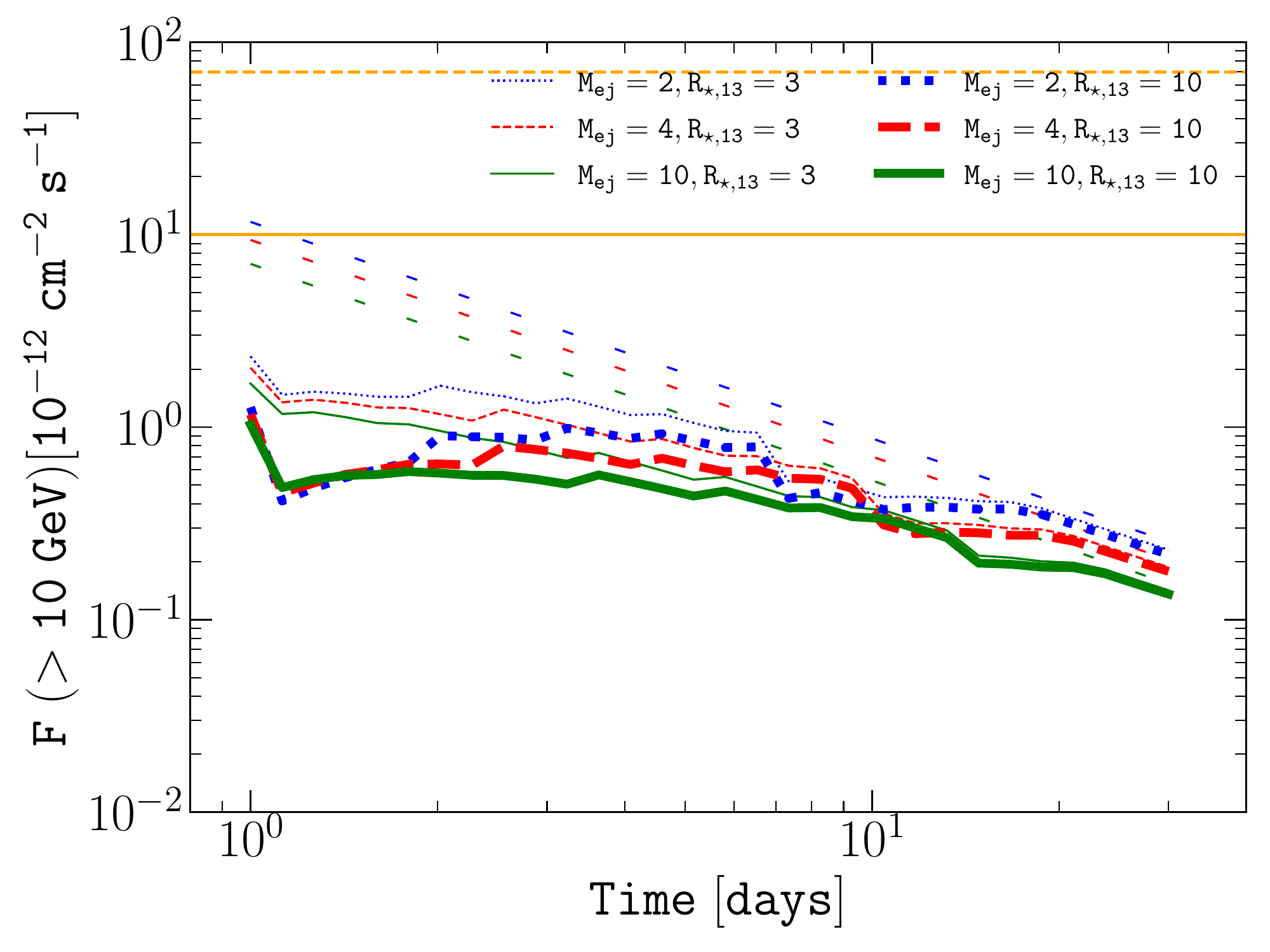}
\includegraphics[width=.48\textwidth]{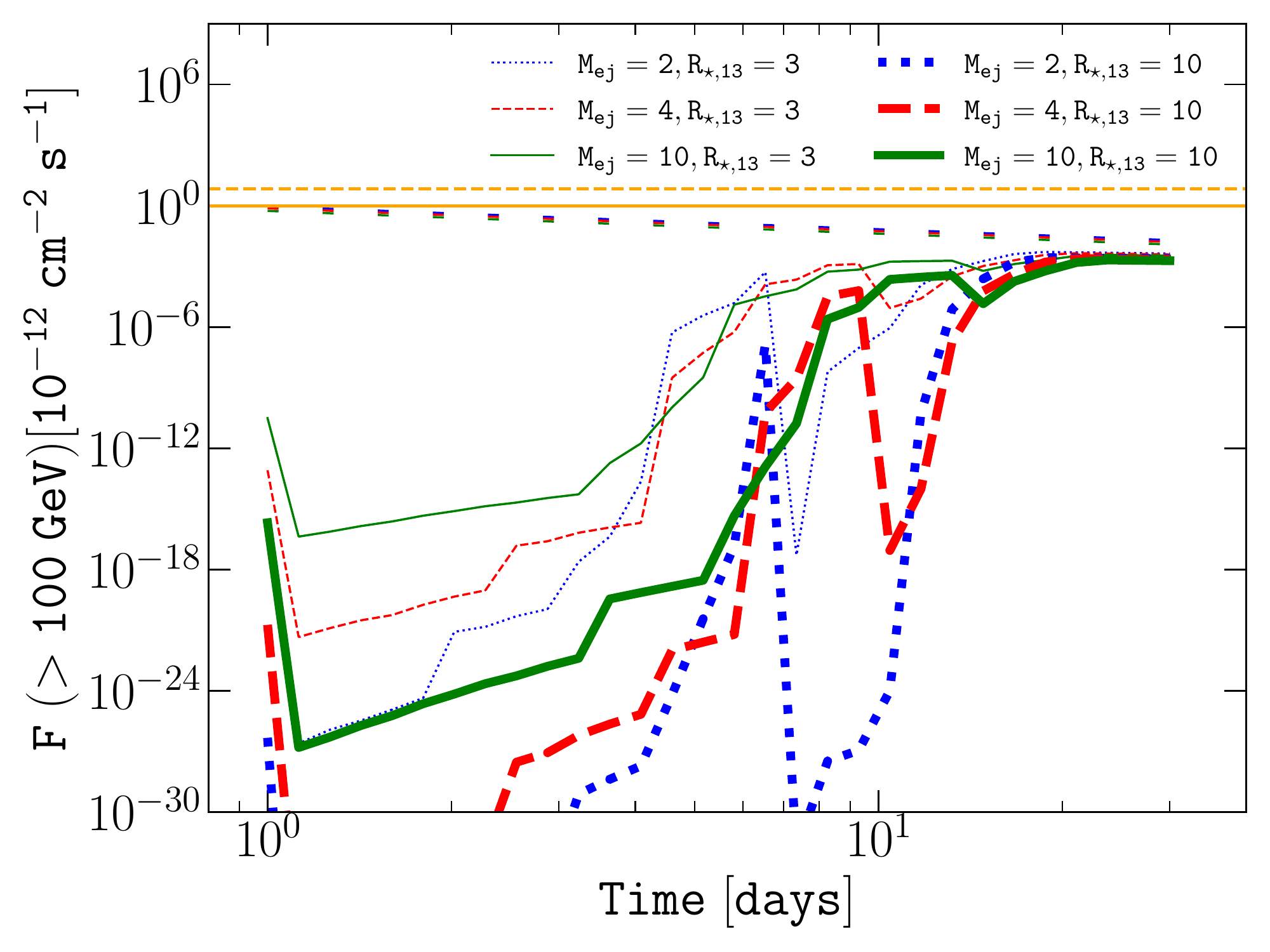}
\includegraphics[width=.48\textwidth]{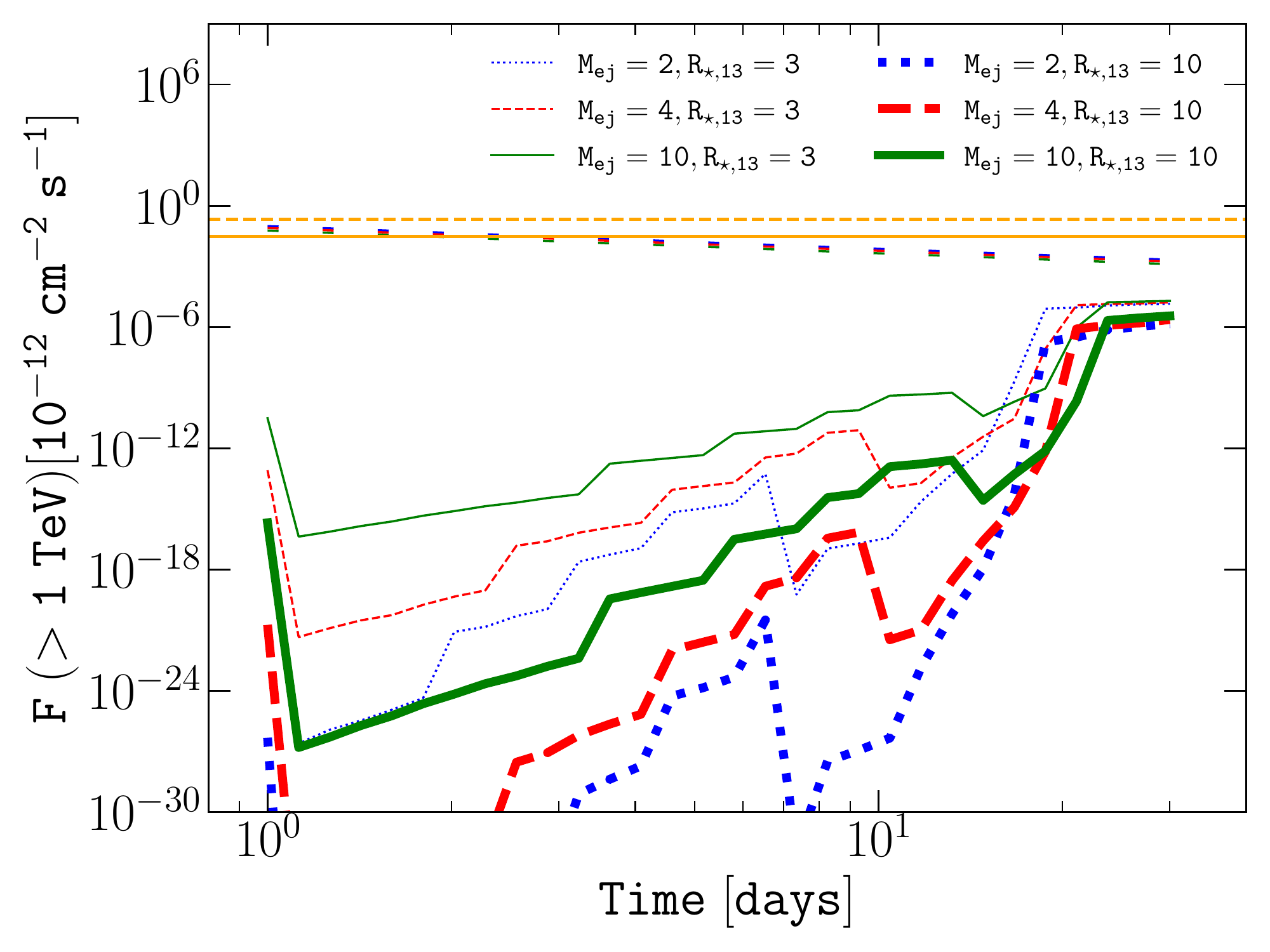} \\
\caption{Time evolution of the integrated photon flux above 10 GeV (top), 100 GeV (middle), and 1 TeV (bottom). The source distance is $D=1$~Mpc, the mass-loss rate of the RSG wind is $\dot{M}_{\rm w}=10^{-6}$ $M_{\odot}$/yr. The results for an ejecta mass $M_{\rm ej}=2, 4$ and 10 M$_{\odot}$ are shown as dotted (blue), dashed (red), solid (green) for a progenitor radius $R_{\star}=3 \times 10^{13}$~cm (thin) and $R_{\star}=10^{14}$~cm (thick) lines. The corresponding unattenuated fluxes are shown as loosely dotted lines. The typical sensitivity of CTA for 50 hours (solid orange  horizontal line) and 2 hours (dashed orange  horizontal line) is shown as a guiding-eye for the reader~\citep{fioretti2016}.}
\label{fig:R}
\end{figure}

In Fig.~\ref{fig:Mej4} and Fig.~\ref{fig:R}, we explore the parameter space and illustrate our results. Fig.~\ref{fig:Mej4} also includes a comparison to prior results. For display purposes, we only show the gamma-ray fluxes obtained under the approximation of an homogeneous source~\citep{wang2019} for fluxes integrated above 100 GeV, where the difference is the greatest. We start by showing, for a fixed ejecta mass $M_{\rm ej}=4$ M$_{\odot}$, the effect of various mass-loss rates. The total explosion energy is kept equal to $E_{\rm SN}=10^{51}$ erg. As written in Eq.~\eqref{eq:gamma_flux}, the unabsorbed gamma-ray flux is expected to scale  as $\dot{M}_{\rm w}^2$. However, the scaling of the gamma-gamma opacity with the mass-loss rate is not straightforward. If $r_{\rm ph}$ and $T_{\rm ph}$ are not functions of $\dot{M}_{\rm w}$, $R_{\rm sh}$, it is scaling as $\propto \dot{M}_{\rm w}^{-0.125}$ (assuming $n=10$). The integrated flux for photons of energy greater than 100 GeV and  1 TeV are shown. Considering photons of energy greater than 500 GeV or 5 TeV lead to a gamma-ray signal of somewhat similar shape to the one above 1 TeV. The plots illustrate the importance of the gamma-gamma attenuation at early time (in the first 10-20 days), as expected, when the photosphere and the shock are the closest, and the photosphere luminosity is the highest. The attenuation is potentially reaching 20 orders of magnitude, and the signal progressively returns to the unattenuated curve.

Figs.~\ref{fig:Mej4} and \ref{fig:R} also exhibit a second dip in the attenuation, at $\approx 10-15$ days, visible on all plots, and especially for the flux above 100 GeV for the
$\dot{M}_{\rm w}=10^{-5}$ M$_{\odot}$/yr case (violet dotted line). This feature can be explained as follows.

The importance of the gamma-gamma process depends essentially on the number density of photons (gamma-ray photons and low energy photons) in interaction, as well as their energy because of the threshold to pair production. Let us for now only consider one sphere emitting photons, for example the photosphere, of radius $R_{\rm ph}(t)$. To reach a point (P) located at a given distance $h$ from the photosphere, the first photons emitted at a time $t_0$ at which the photosphere radius was $R_{\rm ph}(t_0)$ will travel during a time $t_1=h/c$. At later time, more photons keep arriving at (P), emitted at different times $t_0*$ (and travelling for shorter time $t_1*$) since the photosphere radius is expanding, and therefore more photons emitted at various times can reach the point (P). This results, in (P), in an increasing number of photons (and thus an increasing luminosity from photons from the photosphere) as $\alpha (t)$ increases (see Fig.~\ref{fig:schema}), until $\alpha$ reaches a maximum value $\alpha_{\rm max}(t)$ (corresponding to a photon emitted at the minimum $t_0^{\star}$). Therefore, at any distance from the photosphere, the number of photons is increasing until reaching a maximum value: the closer points see a quicker increase, and the farther points see a delayed and slower increase. 

\begin{figure}
\includegraphics[width=.48\textwidth]{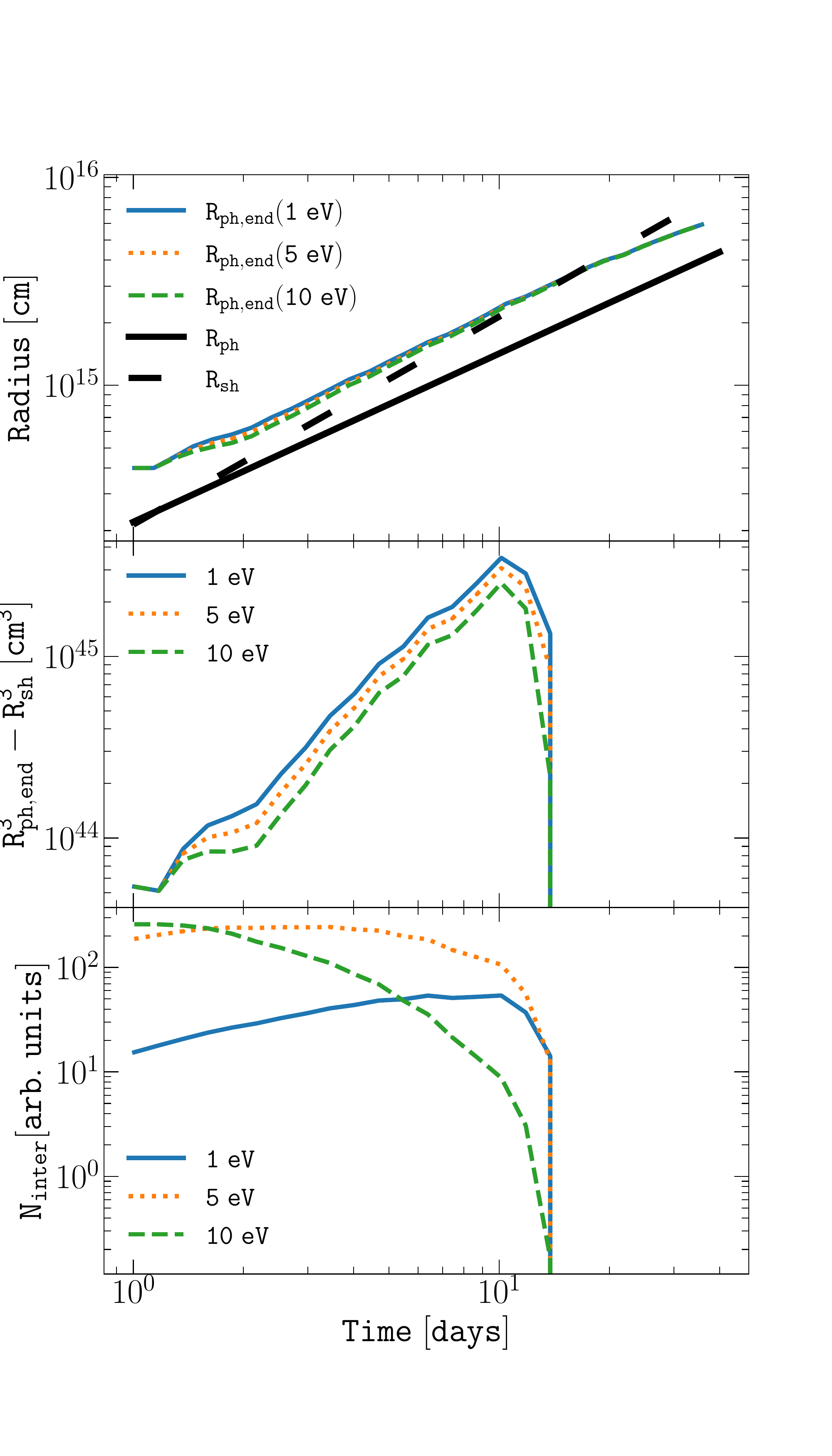}
\caption{Time evolution of $R_{\rm ph,end}$(top panel), of the interaction volume $R_{\rm ph,end}^3- R_{\rm sh}^3$  (middle panel), and of the associated number of photons $N_{\rm inter}$ inside the interaction volume (bottom panel), for photons of energy 1 eV, 5 eV and 10 eV (blue solid, yellow dotted and green dashed): . $E_{\rm SN}=10^{51}$ erg, $M_{\rm ej}=4$ M$_{\odot}$, $\dot{M}=10^{-6}$ M$_{\odot}$/yr,  and $R_{\star}= 10^{14}$ cm. }
\label{fig:multi}
\end{figure}

\begin{figure}
\includegraphics[width=.48\textwidth]{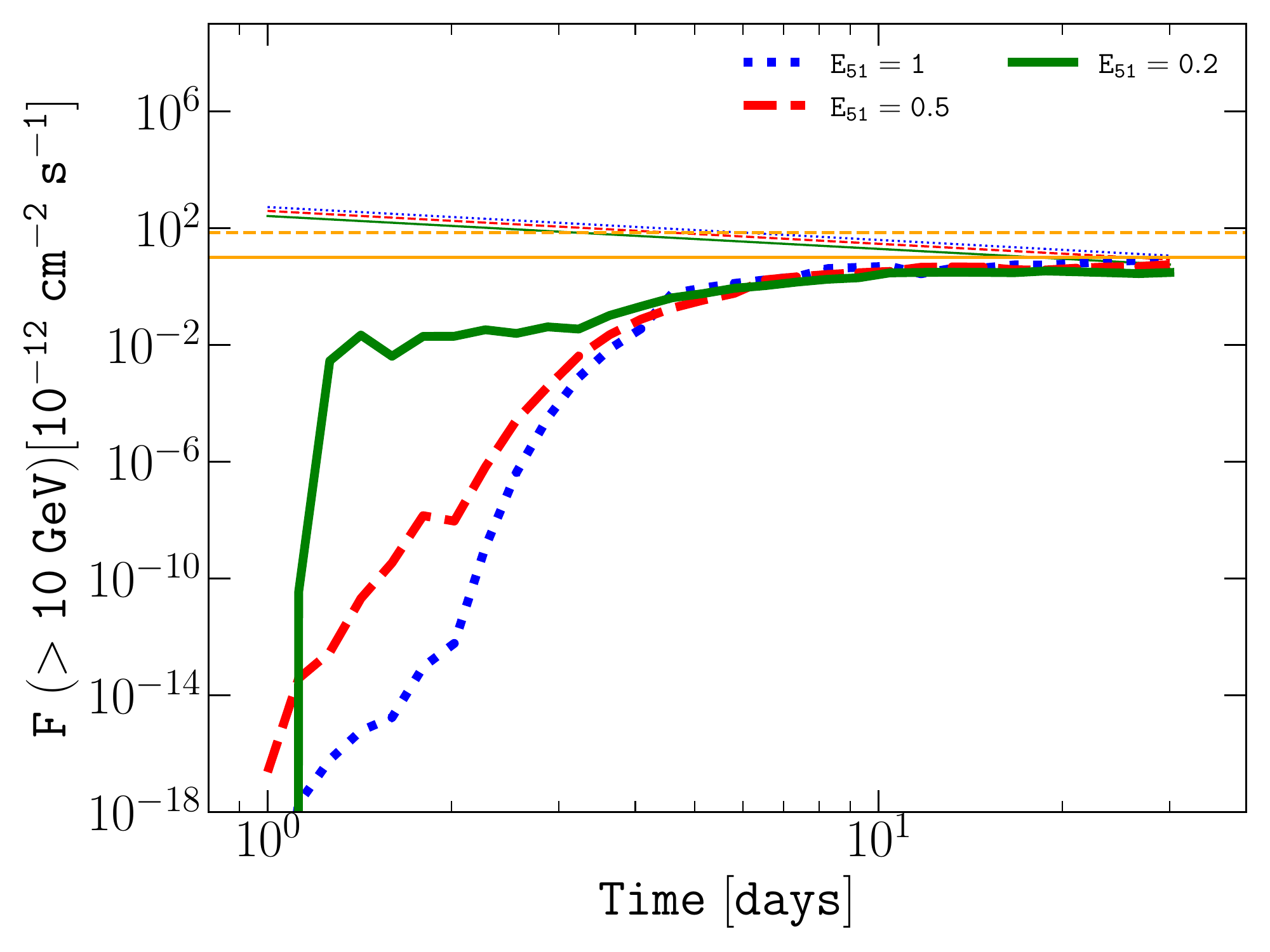}
\includegraphics[width=.48\textwidth]{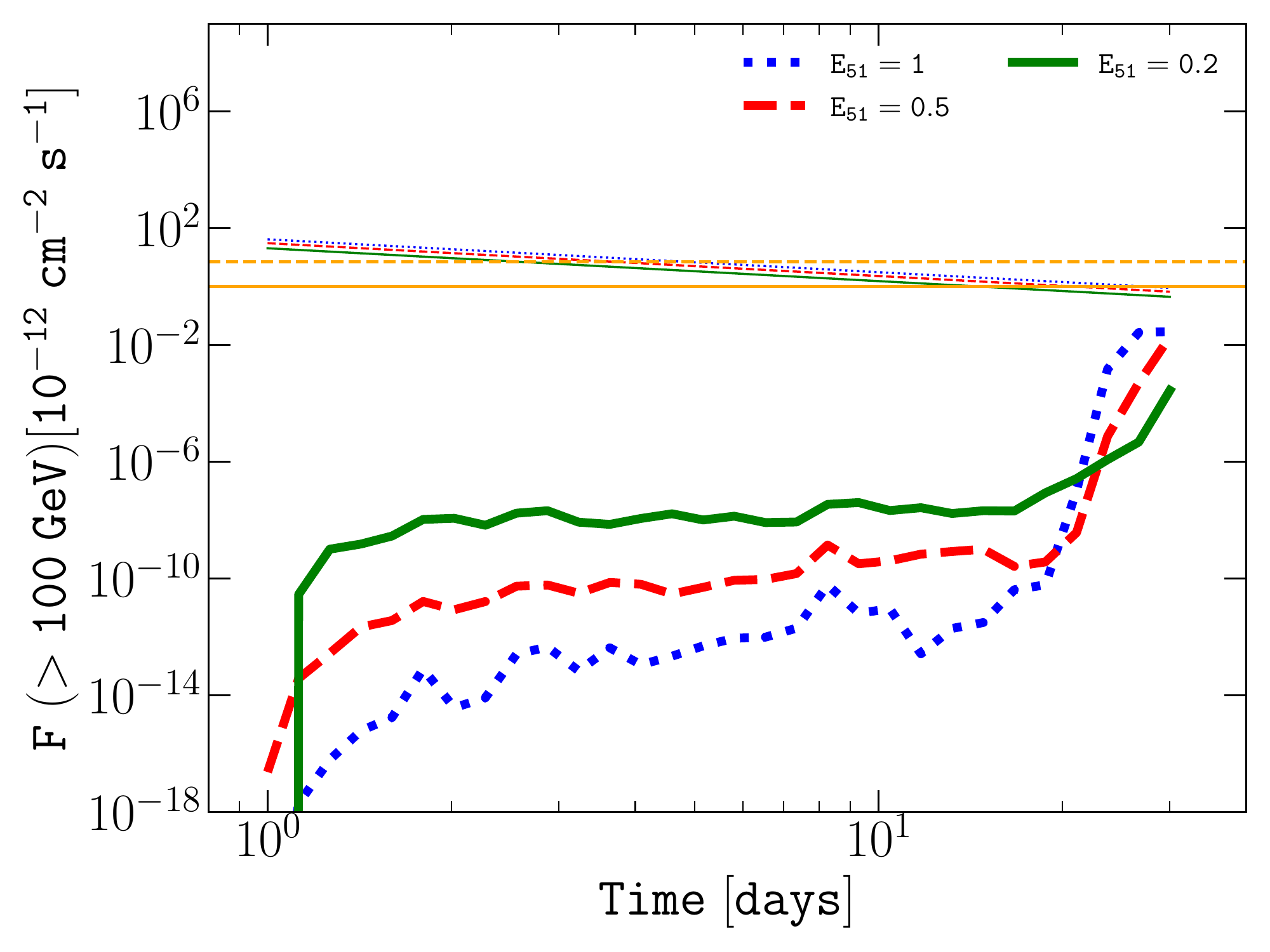}
\includegraphics[width=.48\textwidth]{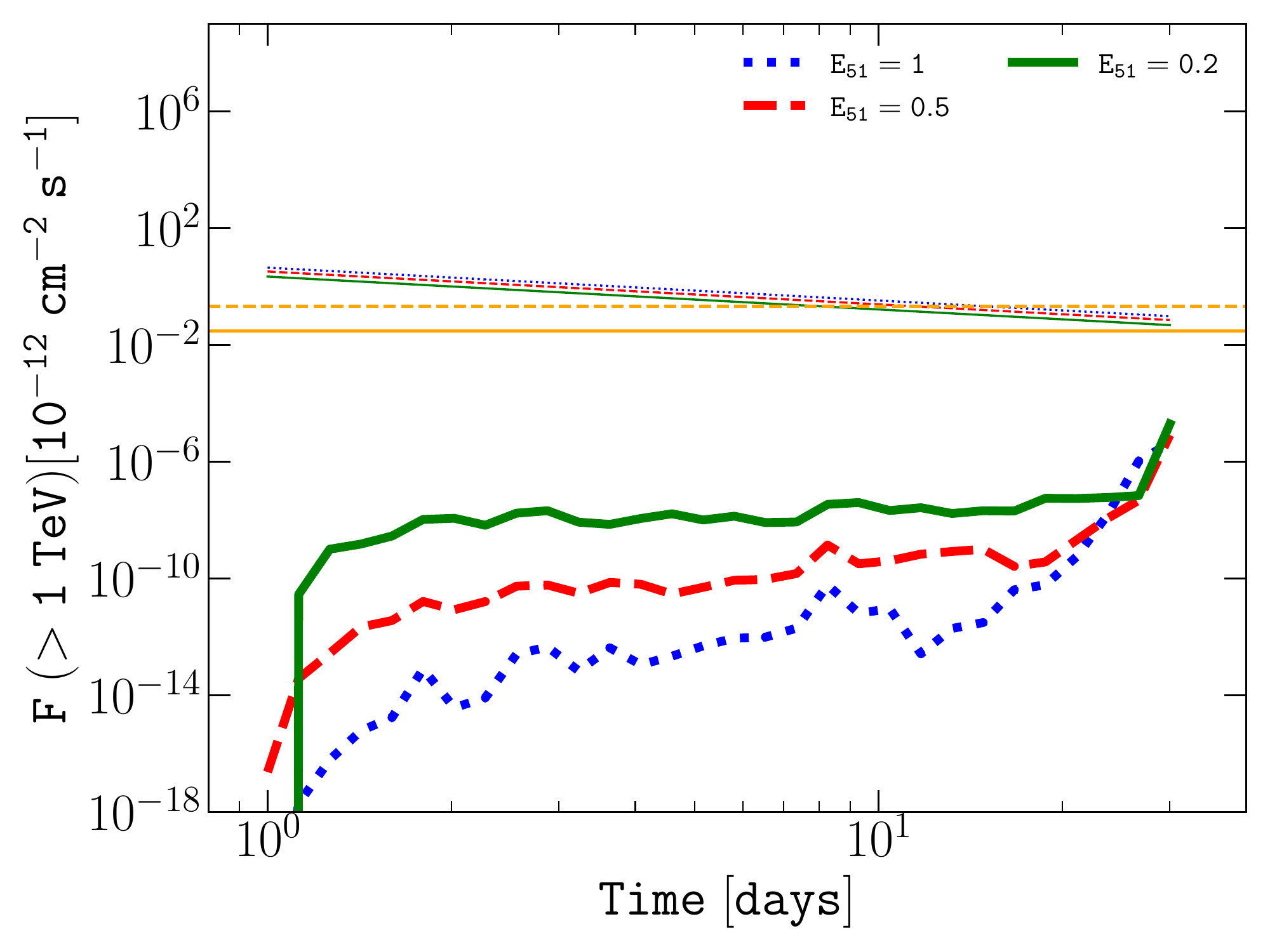}  \\
\caption{Time evolution of the integrated photon flux above 10 GeV (top), 100 GeV (middle) and 1 TeV (bottom). The mass-loss rate of the RSG wind is $\dot{M}_{\rm w}=10^{-5}$ M$_{\odot}$/yr, the radius of the progenitor is $R_{\star}=3 \times 10^{13}$ cm and the ejecta mass $M_{\rm ej}=10$ M$_{\odot}$. The total explosion energy of $E_{\rm SN}=10^{51}$, $0.5\times10^{51}$, and $0.2\times10^{51}$~erg correspond to the green solid, blue dotted, and red dashed curves, respectively. The typical sensitivity of CTA  for 50 hours (solid orange  horizontal line) and 2 hours (dashed orange  horizontal line) is shown as a guiding-eye for the reader~\citep{fioretti2016}.}
\label{fig:E}
\end{figure}

We can try to estimate the typical distance up to which most of the photons from the photosphere are found at a given time $t$. Of course, emitted at $t_0$, a photon can always at most propagate up to $c(t-t_0)$, but as just discussed, the time effect taking into account the delay of photons implies that the bulk of photons will only propagate to a shorter distance, that we name $R_{\rm ph, end}$, which is a function of time. In order to estimate $R_{\rm ph, end}$ at a time $t$, we can calculate the luminosity $L(i,t)$ at different points $i$ located at a distance $h(i)$ from the photosphere: 
\begin{equation}
    R_{\rm ph, end}(t) \approx \frac{\sum_{i=0}^{\infty} h(i) L(i,t)}{\sum_{i=0}^{\infty} L(i,t)}
\end{equation}
We thus evaluate $R_{\rm ph, end}$ that way. For illustrative purposes we plot in Fig.~\ref{fig:multi} $R_{\rm ph}$, $R_{\rm sh}$ and $R_{\rm ph, end}$ for a given set of parameters: $E_{\rm SN}=10^{51}$ erg, $M_{\rm ej}=5$ M$_{\odot}$, $\dot{M}=10^{-5}$ M$_{\odot}$/yr, and $R_{\star}=10^{14}$ cm. 
In order to estimate the number of low energy photons that can interact with gamma rays, we can consider that the interacting photons are located in the volume $V_{\rm inter}(t)\approx 4 \pi/3(R_{\rm ph,end}^3(t)- R_{\rm sh}^3(t))$. This is, of course, a simplified view of the problem, but which illustrates that $V_{\rm inter}(t)$ is non--monotonic and peaking around $\sim 10$ days (see middle panel of Fig.~\ref{fig:multi}). In this volume $V_{\rm inter}(t)$, we can estimate the number of photons $N_{\rm inter}(\epsilon, t)\approx n(\epsilon,t) \epsilon V_{\rm inter}(t)$ (bottom panel of Fig.~\ref{fig:multi}). This calculation illustrates that number of photons in $V_{\rm inter}$ at energy 1 eV close to the peak of interaction with TeV photons \citep{vassiliev2000} increases with time before sharply decreasing at $\sim 10$ days. For photons of slightly different energy, the behavior of $N_{\rm inter}$ with time changes radically: this illustrates, that the number of photons of a given energy can increase after several $\sim 10$ days, and produce the second attenuation dip present in various plots: Figs.~\ref{fig:Mej4} and \ref{fig:R}, and Fig.~\ref{fig:E}. The parameters adopted in Fig.~\ref{fig:multi} correspond exactly to the red thick dashed lines of Fig.~\ref{fig:R}. 
Notice that in the case of SN 1993J~\citep{cristofari2020}, the shock and photosphere are sufficiently distant to ensure that $R_{\rm ph,end}^3- R_{\rm sh}^3$ is monotonous, and the second dip is not visible. 

In Fig.~\ref{fig:R}, we illustrate the effect of a change of temperature (and thus luminosity) of the photosphere by considering two different radii of the progenitor star $R_{\star}=3 \times 10^{13}$ and $10^{14}$~cm. The temporal evolution of the temperature and luminosity is shown in Fig.~\ref{fig:temperature} and the evolution of the black--body distribution is illustrated in Fig.~\ref{fig:photons}. The resulting gamma-ray signal is shown in Fig.~\ref{fig:R}, illustrating that the increased temperature and luminosity lead to an enhancement of the attenuation (visible for all ejecta mass considered). Moreover, the second attenuation dip featured at $\sim 10$ days discussed above is more pronounced as the photosphere luminosity increases (i.e. as the energy available in the overlapping interaction volume $V_{\rm inter}$ increases).
In addition, the effect of the ejecta mass is shown, showing that increasing ejecta mass tends to lower the level of the attenuation (the effect is especially visible for higher temperatures of the photosphere). 

Recent works~\citep[see e.g.][]{barkeretal21} mentioned that for high mass-loss rates, the explosion energy of SNe arising from stars of mass $\lesssim$ 20 $\msun$ can be less than $10^{51}$ erg. Interestingly, it is possible to understand the effect of a lower total explosion energy by looking at Eq.~\eqref{eq:rph}, Eq.~\eqref{eq:Tph} and Eq.~\eqref{eq:Rsh2}. The temperature is very weakly dependent on the total explosion energy $E_{\rm SN}$. The photospheric radius and shock radius scale with $E_{\rm SN}$ in the same way $\propto E_{\rm SN}^{0.41}$. This dependency is therefore analogous to the dependency on the mass of the ejecta $M_{\rm ej}$: $r_{\rm ph}$ and $R_{\rm sh}$ both roughly scale as $\propto M_{\rm ej}^{-0.25}$, and thus is affecting the shock radius and photospheric in the same proportion. In other words, for instance, the effect of decreasing $E_{\rm SN}$ by a factor of 3, would roughly decrease $r_{\rm ph}$ and $R_{\rm sh}$ by a factor $\approx (1/3)^{0.41} \approx 0.65$, equivalent to an increase of the ejecta mass by a factor $\approx 0.65^{-1/0.25} \approx 5.6$. The dependence of the absorption on $E_{\rm SN}$ is illustrated in Fig.~\ref{fig:E}.

\section{Conclusions}
\label{sec:conclusions}
We have estimated the gamma-ray signal emerging from typical type II-P CCSNe in the first month after the SN explosion, taking into account non-isotropic time-dependent attenuation due to pair production. After a few tens of days the gamma-ray emission is similar to the unabsorbed solution. Our results can be summarized as follows.

For typical extragalactic CCSNe, the gamma-ray signal, after inclusion of the two-photon annihilation process, is expected to be significantly lower than the typical sensitivity of Cherenkov instruments for objects located at 1 Mpc. This indicates that mostly nearby type II-P SNe, typically exploding in our Galaxy or in the Magellanic Clouds, are expected to be detectable by next generation instruments, such as CTA, although a detailed calculation using the last CTA sensitivity is necessary before providing a more precise limit. Type II-P SNe make up the major fraction of CCSNe. It is worth noting at this stage that our calculation of the pair production does not account for the possibility to produce saturated pair cascades, i.e. to produce several generations of electron-positron pairs in the radiation field produced by the previous generation. Together with the possibility of gamma-ray emission at the reverse shock, these different generation of pairs can add up some gamma-ray emission, especially in the GeV range. So our calculation has to be seen as conservative. The pair cascade process issue will be addressed in a future work. Finally, type II-P SNe are generally interacting with a relatively diluted medium. A study of SN types that evolve in a higher-density medium (such as some type IIb and IIn SNe) will probably greatly improve chances of detection in the gamma-ray range.

Several authors have found that, in order to fit the initial rise and the peak of the optical light curves of Type II-P SNe, they must have experienced a very high mass-loss rate of up to 1 $\msun$ yr$^{-1}$ in the very last few years before the star collapsed to become a SN  \citep{yaron2017,morozovaetal17,morozovaetal18,forsteretal18,rd19}. These high mass-loss rates in many cases occur in only the last 2-3 years before core-collapse. If we assume that they are due to RSG winds with a wind velocity of typically 10 km s$^{-1}$, then the SN shock would be expected to cross this high density region within a few days or $\lesssim 10$ days, depending on the progenitor activity. 
This enhanced mass loss rate  
could 
delay the onset of particle acceleration if the wind is optically thick. Indeed, particle acceleration is expected to start around SN shock breakout at an optical depth $\tau \approx c/v_{\rm sh}$, when the radiation--mediated shock stalls and is replaced with a collisionless shock. In the case of an optically thick wind, this would occur in the wind at a radius $R_{\rm br} > R_{\star}$, rather than at the surface of the star. Assuming a spherical wind with opacity $\kappa$, and with density $\rho(r) = \dot{M}_{\rm w} / (4 \pi v_{\rm w} r^2)$ between $r=R_{\star}$ and $r=R_{\rm w}$ and negligible beyond, the optical depth is $\tau(r) = \kappa \int_r^{R_{\rm w}} \rho \text{d}r$, and $R_{\rm br}$ then satisfies: $R_{\rm br}^{-1} \approx R_{\rm w}^{-1} + 4 \pi c v_{\rm w} / \kappa \dot{M}_{\rm w} v_{\rm s}$ \citep[see also][]{ChevalierIrwin2011}. For large mass-loss rates, such that $R_{\rm w} < \kappa \dot{M}_{\rm w} v_{\rm s} / 4 \pi c v_{\rm w}$, the shock breakout and the onset of particle acceleration would occur at $R_{\rm br} \approx R_{\rm w}$. For more moderate mass-loss rates, such that $R_{\rm w} > \kappa \dot{M}_{\rm w} v_{\rm s} / 4 \pi c v_{\rm w}$, this would occur at $R_{\rm br} \approx \kappa \dot{M}_{\rm w} v_{\rm s} / 4 \pi c v_{\rm w}$. Assuming that $\kappa$ is dominated by Thomson scattering, i.e. $\kappa = \sigma_{\rm t}/m_{\rm p}$ where $\sigma_{\rm t}$ is the Thomson cross section, one finds: $R_{\rm br} \approx 10^{14}\,{\rm cm} (\dot{M}_{\rm w}/5 \cdot 10^{-4}\,{\rm M}_\odot {\rm yr}^{-1}) (v_{\rm sh}/0.1\,c) (v_{\rm w}/10\,{\rm km \,s^{-1}})^{-1}$. If the wind is optically thick, $R_{\rm br} > R_{\star}$, and particle acceleration can start at $r \gtrsim R_{\rm br}$ when conditions are favourable, see~\citet{Giacinti15} for specific cases where particle acceleration could start at $r < R_{\rm br}$. On the contrary, if $R_{\rm br} < R_{\star}$, the wind is, in fact, optically thin: the shock breakout occurs at $R_{\rm br} = R_{\star}$, and particle acceleration may start at $r \gtrsim R_{\star}$. In case the particle acceleration onset is not much delayed, enhanced mass loss rates should conversely lead to more efficient particle acceleration and possibly enhanced gamma-ray emission. In effect, higher $\dot{M}/v_{\rm w}$ ratios produce higher CR-driven instability growth rates \citep{marcowith2018}. Ultimately, higher magnetic field strengths should be obtained at the shock front. This should produce higher CR energies eventually reaching the CR knee within a week after the shock breakout~\citep{Inoue21}.

SN 1987A, the closest SNe to us in over 300 years, deserves special mention here. Sometimes this is considered similar to the II-P SNe, or treated as Type II$_{\rm peculiar}$. However its progenitor was a blue supergiant, not a RSG~\citep{mccray2016}. The SN shock wave initially evolved in a very low density blue supergiant wind with a mass-loss rate $\dot{M} \sim 10^{-9} - 10^{-8}$ M$_{\odot}$ yr$^{-1}$ \citep{cd95, lundqvist99,deweyetal12}, lower than the SNe considered herein. After a few years, the SN interacted with higher density circumstellar material formed by the blue supergiant wind sweeping up the red supergiant wind from a prior epoch. This interaction could lead to a potentially significant gamma-ray signal \citep{dbk1995,bkv11,dwarkadas2013,bkv15}. It has not been detected by current Imaging Atmospheric Cherenkov Telescopes \citep{hesslmc}, but will be a target for the upcoming CTA. The treatment herein is not applicable to SN 1987A, since the complicated structure of the surrounding medium does not make it amenable to an analytic treatment. 

Let us however note that a typical type II-P SN with a mass loss rate $\gtrsim 10^{-6}$ M$_{\odot}$/yr, located at a distance comparable to the one of SN1987A ($\sim 51$ kpc) would a priori be detectable by next generation instruments such as CTA, this can be seen by rescaling with the distance results of Fig.~\ref{fig:R}.  

The simple parametrization adopted here, for the
shock evolution, and photospheric evolution, allows to study the gamma-gamma absorption and discuss the importance of the physical parameters (mass of the ejecta, mass-loss rate of the RSG wind in which the SN explodes, temperature of the photosphere, total explosion energy, opacity, radius of the exploding star). We especially illustrate how changes by a factor of $\sim 2-3$ on some parameters (e.g., the stellar radius, and thus on the expected temperature) can lead to dramatic changes in the level of absorption. The time-integration of absorption effects can in some situations lead to detectable features in the gamma-ray signal, such as when the shock radius evolves very close to the photosphere (e.g., $\dot{M_{\rm w}} \sim 10^{-5}$ M$_{\odot}$/yr). 
Finally, we illustrate that the largest gamma--ray fluxes from type II-P CCSNe in the first days after the SN explosion, are expected when $\dot{M_{\rm w}}$, $M_{\rm ej}$ are the highest, $E_{\rm SN}$ and $R_{\star}$ are the smallest, and could potentially be detected by next generation instruments, such as CTA~\citep{CTAscience}, the Southern Wide-field Gamma-ray Observatory~\citep{SWGO}, or the Large High Altitude Air Shower Observatory~\citep{LHAASO}.

\section*{Acknowledgements}
VVD is supported by National Science Foundation grant 1911061 awarded
to the University of Chicago (PI: Vikram Dwarkadas). 
The research activity of E.P. was supported by Villum Fonden (project n. 18994) and by the European Union’s Horizon 2020 research and innovation program under the Marie Sklodowska-Curie grant agreement No. 847523 ‘INTERACTIONS’.

 \section*{Data Availability}

No data has been analyzed or produced in this work.



\bibliographystyle{mnras}
\bibliography{CCSNe} 



\bsp	
\label{lastpage}
\end{document}